# Evaluation of graphene optical nonlinearity with photon-pair generation in graphene-on-silicon waveguides

YUYA YONEZU,[1,2,7] RAI KOU,[3,4,5] HIDETAKA NISHI,[3,4] TAI TSUCHIZAWA,[3,4] KOJI YAMADA,[3,4,5] TAKAO AOKI,[2] ATSUSHI ISHIZAWA,[1] AND NOBUYUKI MATSUDA[1,4,6,8]

[1]*NTT Basic Research Laboratories, NTT Corporation, Morinosato-Wakamiya, Atsugi, Kanagawa 243-0198, Japan*
[2]*Department of Applied Physics, Waseda University, Okubo, Shinjuku, Tokyo 169-8555, Japan*
[3]*NTT Device Technology Laboratories, NTT Corporation, Morinosato-Wakamiya, Atsugi, Kanagawa 243-0198, Japan*
[4]*NTT Nanophotonics Center, NTT Corporation, Morinosato-Wakamiya, Atsugi, Kanagawa 243-0198, Japan*
[5]*Currently at National Institute of Advanced Industrial Science and Technology (AIST), Tsukuba, Ibaraki 305-8569, Japan*
[6]*Currently at Department of Communications Engineering, Graduate School of Engineering, Tohoku University, Aramaki Aza Aoba, Aoba-ku, Sendai 980-8579, Japan*
[7]*yonezu@ruri.waseda.jp*
[8]*n.matsuda@ecei.tohoku.ac.jp*

**Abstract:** We evaluate the nonlinear coefficient of graphene-on-silicon waveguides through the coincidence measurement of photon-pairs generated via spontaneous four-wave mixing. We observed the temporal correlation of the photon-pairs from the waveguides over various transfer layouts of graphene sheets. A simple analysis of the experimental results using coupled-wave equations revealed that the atomically-thin graphene sheets enhanced the nonlinearity of silicon waveguides up to ten-fold. The results indicate that the purely $\chi^{(3)}$-based effective nonlinear refractive index of graphene is on the order of $10^{-13}$ m$^2$/W, and provide important insights for applications of graphene-based nonlinear optics in on-chip nanophotonics.

## 1. Introduction

Graphene, a two-dimensional honeycomb nanostructure of carbon atoms, has attracted a great deal of attention in the past couple of decades because of its unique physical properties [1–5]. Of these, graphene's third-order optical nonlinearity, which is responsible for its saturable absorption and nonlinear refractive index change, has been shown to be extremely large due to the linear band structure of $\pi$-electrons [6]. This remarkable property along with the ultrafast carrier dynamics [7] is highly beneficial for the application of graphene to nonlinear-optical processes such as mode-locking of ultrafast pulsed lasers [8–10], wavelength conversion [5,11–13], nonlinear harmonic generation [7,14,15], and supercontinuum generation [16].

It is essential to characterize the material property of graphene for the applications. There are large discrepancies between the reported values of the third-order nonlinear coefficient of graphene. For example, the experimentally obtained effective nonlinear refractive index $n_{2,\text{eff}}$ of graphene ranges from $10^{-11}$–$10^{-13}$ m$^2$/W for stimulated four-wave mixing (FWM) [5,11–13,17] or optical Kerr effect (OKE) [16,18–25] to $10^{-15}$–$10^{-17}$ m$^2$/W for third-harmonic generation (THG) [14,15]. Not only the absolute value, but also the positive/negative sign of the effective nonlinear refractive index $n_{2,\text{eff}}$ of graphene has been under active investigation [23,24]. From theoretical point of view, the 4–6-order difference between the nonlinear coefficients corresponding to the FWM (or the OKE) and the THG has been predicted recently [26–28]. The negative sign of $n_{2,\text{eff}}$ measured in [23] has also been described by a rigorous treatment of the third-order nonlinear conductivity tensor [29]. On the other hand, as for experimental works, the remaining 3-order discrepancy could originate from the material preparation (chemical potential) and experimental methods (see discussions in e.g., [13,23,29–31]). In addition, the contribution of carrier dynamics [7,25] rather than electronic $\chi^{(3)}$ nonlinearity to optical nonlinear phenomena would also lead to the discrepancy. To resolve the discrepancy, further investigations from various viewpoints, e.g., evaluation methods to distinguish between the $\chi^{(3)}$ nonlinearity and the photoexcited-carrier effect, are very important. Note that the simultaneous evaluation of both real and imaginary parts of the nonlinear coefficient using electrically-tunable graphene-covered silicon nitride waveguides [31] is a remarkable example in this context.

Photon coincidence measurement of correlated photon-pairs generated via spontaneous four-wave mixing (SFWM) is a useful method to evaluate the $\chi^{(3)}$ nonlinearity, and is routinely used for quantum optical experiments [32–37]. The coincidence measurement has two key advantages. First, it allows us to easily preclude photon generation from unwanted events, such as Raman scattering. Second, it can be conducted with a simple experimental setup consisting of a pump laser source, passive filters, and single-photon detectors. Hence, it enables us to validly estimate the $\chi^{(3)}$ nonlinearity of a sample simply from the pump pulse information and the total photon loss of the system.

In this paper, we evaluate the nonlinear coefficient of graphene-on-silicon waveguides (GSWs) through the coincidence measurement of photon-pair generation via SFWM. We have observed the temporal correlation of the generated photon-pairs from GSWs over various transfer layouts (lengths and positions) of graphene sheets on silicon waveguides. To the best of our knowledge, this is the first observation of correlated photon-pairs from graphene. According to a simple theoretical analysis using coupled-wave equations [38–41], up to a ten-fold graphene-induced optical nonlinearity enhancement (NLE) was estimated from the observed photon-pair generation rates. The NLE agrees well with our previously reported value [16] estimated via supercontinuum generation in GSWs, which were fabricated with the same process as the samples in this work.

## 2. Device description

Figure 1 shows a schematic of our GSW [16]. The silicon waveguides, with the fixed total length $L$ of 0.5 mm and the widths $W$ of 500, 600, and 700 nm, were fabricated on a silicon-on-insulator (SOI) substrate with a 250-nm-thick silicon layer using electron-beam (EB) lithography and electron cyclotron resonance-reactive ion etching (ECR-RIE). Note that, in this work, we chose shorter waveguides than those used in [16] ($L$ = 2 mm) to explicitly show the contribution of graphene. A monolayer graphene sheet (p-doped, the Fermi energy of 0.2–0.35 eV [42]) was transferred onto the waveguides and patterned using photolithography and $O_2$-plasma reactive ion etching (RIE). The graphene width was fixed at 30 $\mu$m and the graphene lengths $L_{\text{gr}}$ were 10, 30, 50, 100, and 200 $\mu$m. Three graphene positions were investigated: (i) centered ($L_{\text{in}} = (L - L_{\text{gr}})/2$), (ii) fixed input-length I ($L_{\text{in}}$ = 0.3 mm), and (iii) fixed input-length II ($L_{\text{in}}$ = 0.15 mm). The measured propagation losses $\alpha_{\text{gsw}}$ of the graphene-covered region were 0.111±0.006, 0.081±0.004, and 0.067±0.005 dB/$\mu$m for $W$ of 500, 600, and 700 nm, respectively. Spot-size converters (SSCs) with $SiO_2$ over cladding (OVC) were also fabricated on both sides of the GSWs for low-loss coupling with lensed fibers [43]. The SSCs have a tapered length of 0.15 mm and a tip width of 200 nm.

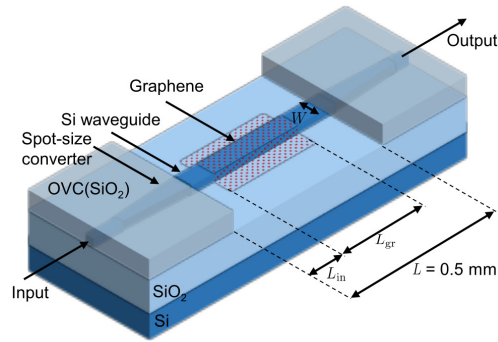

**Fig. 1.** Graphene-on-silicon waveguide (GSW). $L$ is the total waveguide length and $W$ is the waveguide width. $L_{gr}$ and $L_{in}$ are the lengths of graphene-covered region and input-side silicon waveguide, respectively. Spot-size converters (SSCs) with $SiO_2$ over cladding (OVC) were fabricated on both sides of the GSWs for low-loss coupling using lensed fibers.

## 3. Experimental methods

Figure 2 shows a schematic of the experimental setup. A continuous-wave (CW) pump laser operated at a wavelength $\lambda_p$ of 1551.1 nm was modulated into pulses with a duration $\Delta\tau$ of 20 ps and a repetition rate $R_p$ of 1 GHz by a $LiNbO_3$ intensity modulator (IM). The pulses were amplified by an erbium-doped fiber amplifier (EDFA), whose amplified spontaneous emission (ASE) noise was reduced by using a band-pass filter (BPF). Then, the pulses were injected into the TE-like fundamental guided mode of the GSWs with a low-loss coupling ($\eta_{couple} \sim -1.4$ dB/facet) between a lensed fiber and the SSC [43].

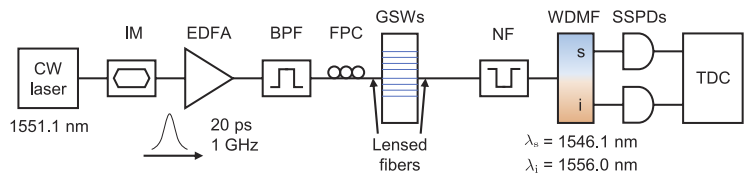

**Fig. 2.** Experimental setup for the photon-pair generation via SFWM in GSWs. IM: $LiNbO_3$ intensity modulator. EDFA: erbium-doped fiber amplifier. BPF: band-pass filter. FPC: fiber polarization controller. NF: notch filter. WDMF: wavelength-division multiplexing filter. SSPD: superconducting single-photon detector. TDC: time-to-digital converter.

To observe the correlated photon-pair generation via SFWM in GSWs, we conducted the coincidence measurement. The output fields were collected into a single-mode fiber by using another lensed fiber. The residual pump pulses were eliminated by using a notch filter (NF). Signal and idler photons were separated into the two output ports of a wavelength-division multiplexing filter (WDMF). Their center wavelengths were 1546.1 nm (signal wavelength $\lambda_s$) and 1556.0 nm (idler wavelength $\lambda_i$) with the bandwidth $\Delta\nu$ of 0.12 THz (0.96 nm). The photons were detected with superconducting single-photon detectors (SSPDs), and their time correlation was measured with a time-to-digital converter (TDC). The total measurement loss $\eta_{tot} = \eta_{couple}\eta_{filter}\eta_{det}$ of our experimental setup was −8.5 dB, where the filter loss $\eta_{filter}$ and the SSPD system detection efficiency $\eta_{det}$ were −6.1 and −1.0 dB (79.6%), respectively. The dark count rate of the SSPDs was 26 Hz.

## 4. Photon-pair generation in graphene-on-silicon waveguides

Figures 3(a)–3(i) show the net photon-pair generation rate at the output end of the GSWs as a function of graphene length $L_{gr}$ with fixed pump peak power $P_p$ of 141 mW. The measurement time for each data point was 60 s. Two GSWs (with four bare silicon waveguides as references) were measured for each waveguide width $W$, graphene length $L_{gr}$, and graphene position. Here, we defined the net photon-pair generation rate per pulse at the output end of the GSWs, $\mu$, as [34,35]

$$\mu = \frac{C_c - C_{c,a}}{R_p \eta_{tot}^2} \quad (1)$$

where $C_c$ and $C_{c,a}$ are the raw coincidence rate (including the accidental coincidence count) and the raw accidental coincidence rate, respectively.

To numerically evaluate the experimental results, we take the standard approach using the coupled-wave equations for the pump, signal, and idler fields propagating along the waveguide (defined as $z$-direction) [38–41]

$$\frac{\partial A_p}{\partial z} + \frac{1}{2}\alpha(z)A_p = i\beta(\omega_p)A_p + i\gamma(z)|A_p|^2 A_p \quad (2)$$

$$\frac{\partial \hat{A}_s(z,\omega_s)}{\partial z} + \frac{1}{2}\alpha(z)\hat{A}_s(z,\omega_s) = i\beta(\omega_s)\hat{A}_s(z,\omega_s) \\ + i2\gamma(z)|A_p|^2\hat{A}_s(z,\omega_s) + i\gamma(z)A_p^2 \hat{A}_i^\dagger(z,\omega_i) \quad (3)$$

$$\frac{\partial \hat{A}_i^\dagger(z,\omega_i)}{\partial z} + \frac{1}{2}\alpha(z)\hat{A}_i^\dagger(z,\omega_i) = -i\beta(\omega_i)\hat{A}_i^\dagger(z,\omega_i) \\ - i2\gamma(z)|A_p|^2 \hat{A}_i^\dagger(z,\omega_i) - i\gamma(z)A_p^{*2}\hat{A}_s(z,\omega_s) \quad (4)$$

where $\beta(\omega_j)$ (j = p, s, i), $\alpha(z)$, and $\gamma(z)$ denote propagation constants at frequency $\omega_j$, propagation loss, and a nonlinear coefficient, respectively. For simplicity, we assume that the propagation loss $\alpha(z)$ is the same for all the fields and that the propagation constants $\beta(\omega_j)$ of the graphene-covered region are the same as that of the bare silicon waveguide region. Note that the $z$-dependence of the propagation loss $\alpha(z)$ and the nonlinear coefficient $\gamma(z)$ are explicitly included in our calculation to investigate the graphene-induced loss and nonlinearity enhancement. The intense pump field amplitude $A_p$ can be treated classically, and the signal and idler field operators $\hat{A}_s(z,\omega_s)$ and $\hat{A}_i(z,\omega_i)$ are treated quantum mechanically in the Heisenberg picture. The pump field amplitude $A_p$ is normalized so that the pump peak power $P_p$ satisfies $P_p = |A_p|^2$. The field operators are normalized by the commutation relation $[\hat{A}_u(z,\omega_u), \hat{A}_v^\dagger(z,\omega_v)] = 2\pi\delta(\omega_u - \omega_v)$. Cross nonlinearities induced by signal and idler fields are neglected.

In our experimental conditions, we assume that two-photon absorption (TPA) loss and free-carrier absorption (FCA) loss are negligibly small due to the relatively low pump power [40] and that the phase-matching condition $\Delta\beta = 2\beta(\omega_p) - \beta(\omega_s) - \beta(\omega_i) = 0$ is satisfied because of the small frequency detuning between the pump and signal/idler and the short waveguide length $L$. With these two assumptions, the above coupled-wave equations can be solved analytically and the photon-pair generation rate via the SFWM is given by

$$\mu_{theory} = \Delta\nu\Delta\tau G(L)^2 \eta(L)^2 \quad (5)$$

where

$$G(L) = \int_0^L \gamma(z')P_p \eta(z')dz' \quad (6)$$

$$\eta(L) = \exp\left[-\int_0^L \alpha(z')dz'\right] \quad (7)$$

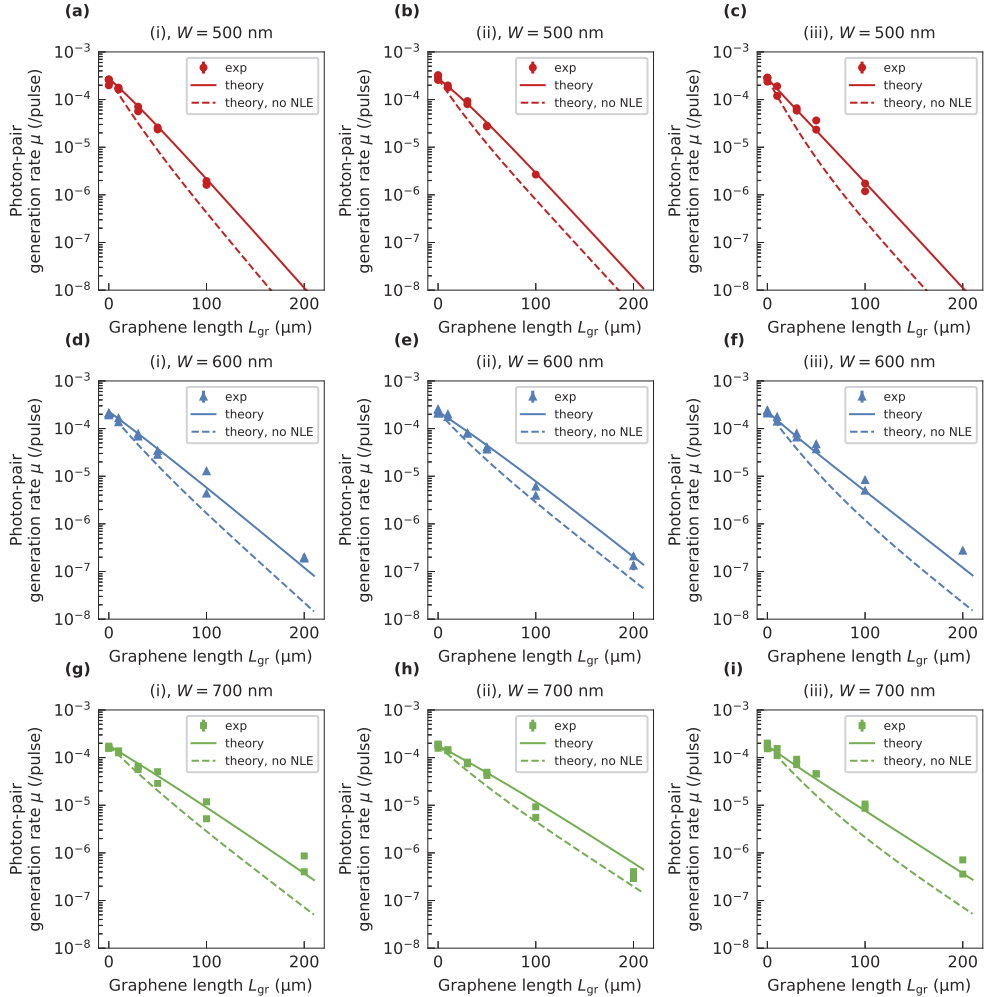

**Fig. 3.** Photon-pair generation rate as a function of graphene length $L_{gr}$ with fixed pump peak power $P_p$ of 141 mW. The data points at $L_{gr} = 0$ represents the case of the bare silicon waveguides as references. (a)–(c) The waveguide width $W$ is fixed at 500 nm. The graphene positions are (i) centered ($L_{in} = (L - L_{gr})/2$), (ii) fixed input-length I ($L_{in} = 0.3$ mm), and (iii) fixed input-length II ($L_{in} = 0.15$ mm), respectively. (d)–(f), (g)–(i) Same as (a)–(c), for $W$ of 600 and 700 nm, respectively. The solid lines represent the results of theoretical analysis, as described in the text. For comparison, the dashed lines show the case without graphene-induced optical NLE (i.e., $\gamma_{gsw} \to \gamma_{sw}$, while the other parameters, e.g., $\alpha_{sw}$, $\alpha_{gsw}$, $L_{gr}$, and $L_{in}$, are the same as the solid lines to include the contribution of the graphene-induced loss). Error bars (smaller than the symbol size in most cases) are calculated assuming the Poisson statistics. Note that the theoretical curves depend on the graphene position as indicated in Eq. (8).

Note that, for a uniform waveguide (i.e., $\alpha(z) = \alpha_0$ and $\gamma(z) = \gamma_0$), Eq. (5) becomes $\mu_{\text{theory}} = \Delta\nu\Delta\tau(\gamma_0 P_p L_{\text{eff}})^2 \eta_0^2$ (where $L_{\text{eff}} = [1 - \exp(-\alpha_0 L)]/\alpha_0$ is the effective waveguide length and $\eta_0 = \exp(-\alpha_0 L)$ is the linear loss factor), which is consistent with the results of previous work [40,44].

For our GSWs, Eqs. (6) and (7) can be rewritten as

$$G(L) = P_p \left[ \gamma_{\text{sw}} \frac{1 - e^{-\alpha_{\text{sw}} L_{\text{in}}}}{\alpha_{\text{sw}}} \right.$$
$$+ \gamma_{\text{gsw}} e^{-\alpha_{\text{sw}} L_{\text{in}}} \frac{1 - e^{-\alpha_{\text{gsw}} L_{\text{gr}}}}{\alpha_{\text{gsw}}} \quad (8)$$
$$\left. + \gamma_{\text{sw}} e^{-\alpha_{\text{sw}} L_{\text{in}}} e^{-\alpha_{\text{gsw}} L_{\text{gr}}} \frac{1 - e^{-\alpha_{\text{sw}}(L - L_{\text{gr}} - L_{\text{in}})}}{\alpha_{\text{sw}}} \right]$$

$$\eta(L) = e^{-\alpha_{\text{sw}}(L - L_{\text{gr}})} e^{-\alpha_{\text{gsw}} L_{\text{gr}}} \quad (9)$$

where $\alpha_{\text{sw}}$ ($\alpha_{\text{gsw}}$) and $\gamma_{\text{sw}}$ ($\gamma_{\text{gsw}}$) are, respectively, the propagation loss and nonlinear coefficient of the bare silicon waveguide region (the graphene-covered region). Equation (8) consists of three terms: the contributions of SFWM in the input-side silicon waveguide, the graphene-covered region, and the output-side silicon waveguide.

By curve fitting of the experimental results for each waveguide width $W$ and graphene position using Eqs. (5), (8), and (9), we estimated the nonlinear coefficients $\gamma_{\text{sw}}$ and $\gamma_{\text{gsw}}$ with the fixed propagation losses $\alpha_{\text{sw}} = 2.0$ dB/cm and $\alpha_{\text{gsw}} = 0.111, 0.081$, and $0.067$ dB/$\mu$m for $W$ of 500, 600, and 700 nm, respectively. The average value of the measured nonlinear coefficients $\gamma_{\text{sw}}$ and $\gamma_{\text{gsw}}$ as a function of the waveguide width $W$ are summarized in Table 1. We obtained a six to tenfold graphene-induced optical NLE for all the waveguide width $W$. When waveguide width $W$ increases, the nonlinear coefficients $\gamma_{\text{gsw}}$ of the graphene-covered region decrease due to decreasing electric field intensity at the graphene positions in the GSWs. The solid lines in Figs. 3(a)–3(i) represent the theoretical curve with the average value of the measured nonlinear coefficients, whereas the dashed lines show the case without graphene-induced optical NLE (i.e., $\gamma_{\text{gsw}} \rightarrow \gamma_{\text{sw}}$, while the other parameters, e.g., $\alpha_{\text{sw}}$, $\alpha_{\text{gsw}}$, $L_{\text{gr}}$, and $L_{\text{in}}$, are the same as the solid lines to include the contribution of the graphene-induced loss). Although the photon-pair generation rate $\mu$ monotonically decreases with increasing graphene length $L_{\text{gr}}$ due to the large graphene-induced propagation loss in both cases, the decreasing slopes of the solid lines are gentler than those of the dashed lines (no NLE). Thus, our results clearly reflect the graphene-induced optical NLE.

Table 1. Measured nonlinear coefficients $\gamma_{\text{sw}}$ and $\gamma_{\text{gsw}}$ as a function of waveguide width $W$.

| $W$ (nm) | $\gamma_{\text{sw}}$ (/W/m) | $\gamma_{\text{gsw}}$ (/W/m) | $\gamma_{\text{gsw}}/\gamma_{\text{sw}}$ |
|---|---|---|---|
| 500 | 158±6 | 1542±41 | 9.8±0.6 |
| 600 | 143±3 | 895±179 | 6.3±1.2 |
| 700 | 127±2 | 696±180 | 5.5±1.4 |

The NLE obtained here agrees well with our previously reported value [16]. In that work, the nonlinear coefficients $\gamma_{\text{sw}}$ and $\gamma_{\text{gsw}}$ for the waveguide width $W$ of 600 nm were numerically calculated to be 117 and 1150 /W/m, respectively, with the overlap between the mode field distribution and graphene, under the assumption that the effective nonlinear refractive index $n_{2,\text{eff}}$ of graphene is $1 \times 10^{-13}$ m$^2$/W (the nonlinear refractive index $n_2$ of bulk silicon was assumed to be $4.2 \times 10^{-18}$ m$^2$/W [45]). The NLE was confirmed through supercontinuum generation in the GSWs with ultra-short (80 fs) pump pulses. Because this pump timescale was much shorter than the effective free carrier lifetime in GSWs [24,25], the NLE would predominantly originate from the $\chi^{(3)}$ nonlinearity rather than the carrier dynamics. Thus, the agreement with the NLE evaluated through the coherent $\chi^{(3)}$-based nonlinear process (SFWM) in this work would be reasonable. In addition, the NLE is also consistent with the results of several other previous works, such as the stimulated FWM [5,11,12] and femtosecond-pulse experiments of the OKE [22,23,25]. Therefore, although the differences in the material preparation in these previous works should be taken into account, our results indicate that the purely $\chi^{(3)}$-based effective nonlinear refractive index $n_{2,\text{eff}}$ of graphene is on the order of $10^{-13}$ m$^2$/W. Note that, in this work, we use the effective nonlinear refractive index $n_{2,\text{eff}}$ for convenience to compare our results with the various previous works although the conventional nonlinear refractive index $n_2$ for three-dimensional materials cannot be defined for two-dimensional materials, as discussed in several theoretical papers [28,29,46]. As with the recent experimental works [13,31], the use of the two-dimensional nonlinear conductivity (or the two-dimensional nonlinear susceptibility defined in [46]) would be appropriate for future work.

Figure 4(a) shows the net photon-pair generation rate at the output end of the GSWs as a function of pump peak power $P_p$ with the fixed graphene length $L_{gr}$ of 100 $\mu$m and the graphene position of (iii) fixed input-length II ($L_{in} = 0.15$ mm). The solid lines represent the results of curve fitting using Eqs. (5), (8), and (9). Our results clearly reveal the $P_p^2$-dependence of the photon-pair generation rate. This $P_p^2$-dependence validates our theoretical model and indicates that the photon-pairs measured here are generated via SFWM.

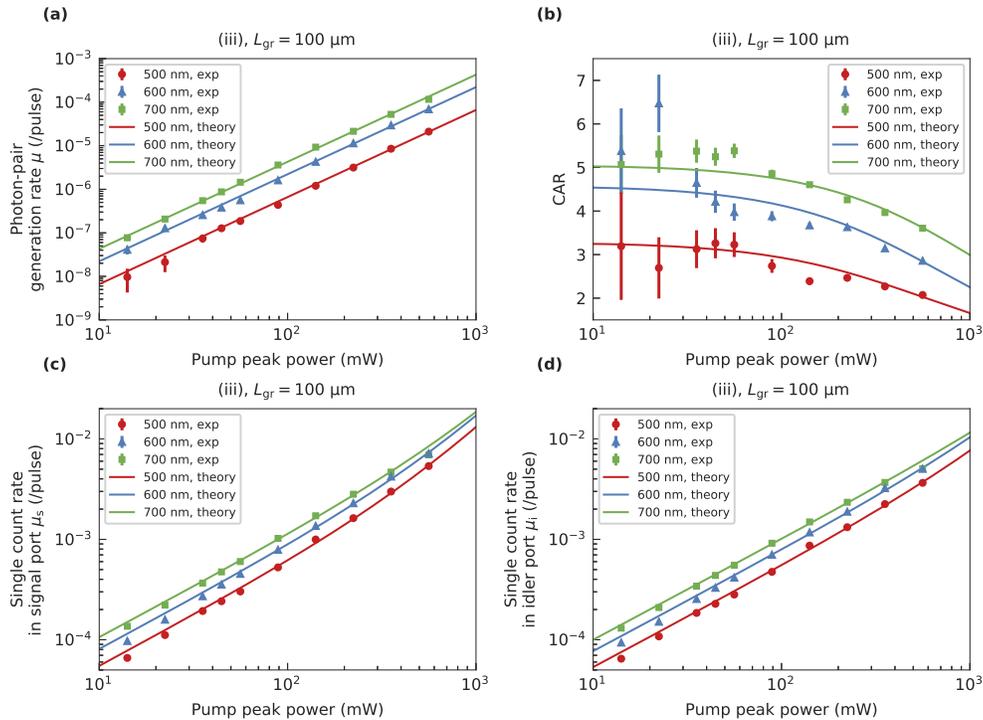

**Fig. 4.** (a) Photon-pair generation rate, (b) coincidence to accidental coincidence ratio (CAR), and (c), (d) single-count rate at the signal and idler ports, as a function of pump peak power $P_p$ with fixed graphene length $L_{gr}$ of 100 $\mu$m and the graphene position of (iii) fixed input-length II ($L_{in} = 0.15$ mm). The solid lines represent the results of theoretical curve fitting described in the text. Error bars are calculated assuming the Poisson statistics.

Finally, Fig. 4(b) shows the measured coincidence to accidental coincidence ratio (CAR) $C_c/C_{c,a}$ as a function of pump peak power $P_p$ with the fixed graphene length $L_{gr}$ of 100 $\mu$m and the graphene position of (iii) fixed input-length II ($L_{in}$ = 0.15 mm). The CAR values of the GSWs (CAR~3–5) are lower than those of the bare silicon waveguides (CAR~20), and decrease with increasing pump peak power. To estimate the CAR values, we define CAR as [34,35]

$$\text{CAR} = 1 + \frac{\mu \eta_{\text{tot}}^2}{(\mu_s \eta_{\text{tot}} + \mu_d)(\mu_i \eta_{\text{tot}} + \mu_d)} \qquad (10)$$

where $\mu_j = C_{s,j}/R_p \eta_{\text{tot}}$ (j = s, i) are the net single-count rates per pulse at the signal/idler ports with the raw single-count rates $C_{s,j}$ (excluding the dark counts) shown in Figs. 4(c) and 4(d). $\mu_d$ denotes the dark count rate per pulse of the SSPDs. The estimated CAR values are shown in Fig. 4(b) by solid lines, and they are in good agreement with the experimental data. In this calculation of the CAR values, we used the fitting curves of $\mu$ in Fig. 4(a) explained above and second-order polynominal fitting curves of $\mu_s$ and $\mu_i$ in Figs. 4(c) and 4(d), respectively. The fitting curves in Figs. 4(c) and 4(d) exhibit the $P_p^1$-dependence of the single-count rates rather than the $P_p^2$-dependence associated with SFWM. This $P_p^1$-dependence indicates that noise photons are generated via processes other than SFWM (e.g., SPM [18,21,24,25] and inelastic scattering [47]) and the CAR of the GSWs has been reduced due to the graphene-induced loss and such unwanted photon-generation events.

## 5. Conclusion

In conclusion, we have evaluated the nonlinear coefficients of GSWs through the coincidence measurement of the photon-pair generation via SFWM. According to the theoretical analysis using the coupled-wave equations [38–41], our results clearly reflected a six to tenfold graphene-induced optical NLE, which is consistent with previously reported values [5,11,12,16,22,23,25]. Because unwanted effects (e.g. Raman scattering) are naturally precluded in our measurement, this consistency indicates that graphene's large optical nonlinearity ($n_{2,\text{eff}} \sim 10^{-13}$ m$^2$/W) purely originates from the $\chi^{(3)}$ nonlinearity rather than the carrier dynamics. Moreover, the photon coincidence measurement method can be easily applied to other two-dimensional materials (BN, MoS$_2$, WS$_2$,...) on photonic waveguides of other materials (SiO$_2$, Si$_3$N$_4$, InP,...). Thus, our results provide helpful insights for an in-depth understanding of the nonlinear optical response of graphene (two-dimensional materials) towards high-performance on-chip nonlinear optical devices.

## Funding

Japan Society for the Promotion of Science (JP17H02803, JP19H02156, JP19K15054, JP26706021); Ministry of Education, Culture, Sports, Science and Technology (A12621600).

## Acknowledgments

This work was supported by JSPS KAKENHI Grant Numbers 17H02803, 19H02156, 19K15054, and 26706021. R. K. acknowledges MEXT Leading Initiative for Excellent Young Researchers (LEADERs) program. Y.Y. acknowledges the Leading Graduate Program in Science and Engineering, Waseda University A12621600 from the Ministry of Education, Culture, Sports, Science and Technology (MEXT), Japan.

## References

1. F. Bonaccorso, Z. Sun, T. Hasan, and A. C. Ferrari, "Graphene photonics and optoelectronics," Nat. Photonics **4**(9), 611–622 (2010).

2. K. Kim, J.-Y. Choi, T. Kim, S.-H. Cho, and H.-J. Chung, "A role for graphene in silicon-based semiconductor devices," Nature **479**(7373), 338–344 (2011).
3. K. S. Novoselov, V. I. Fal'ko, L. Colombo, P. R. Gellert, M. G. Schwab, and K. Kim, "A roadmap for graphene," Nature **490**(7419), 192–200 (2012).
4. Q. Bao and K. P. Loh, "Graphene Photonics, Plasmonics, and Broadband Optoelectronic Devices," ACS Nano **6**(5), 3677–3694 (2012).
5. T. Gu, N. Petrone, J. F. McMillan, A. van der Zande, M. Yu, G.-Q. Lo, D.-L. Kwong, J. Hone, and C. W. Wong, "Regenerative oscillation and four-wave mixing in graphene optoelectronics," Nat. Photonics **6**(8), 554–559 (2012).
6. E. Dremetsika and P. Kockaert, "Enhanced optical Kerr effect method for a detailed characterization of the third-order nonlinearity of two-dimensional materials applied to graphene," Phys. Rev. B **96**(23), 235422 (2017).
7. M. Baudisch, A. Marini, J. D. Cox, T. Zhu, F. Silva, S. Teichmann, M. Massicotte, F. Koppens, L. S. Levitov, F. J. G. de Abajo, and J. Biegert, "Ultrafast nonlinear optical response of Dirac fermions in graphene," Nat. Commun. **9**(1), 1018 (2018).
8. Q. Bao, H. Zhang, Y. Wang, Z. Ni, Y. Yan, Z. X. Shen, K. P. Loh, and D. Y. Tang, "Atomic-Layer Graphene as a Saturable Absorber for Ultrafast Pulsed Lasers," Adv. Funct. Mater. **19**(19), 3077–3083 (2009).
9. I. H. Baek, H. W. Lee, S. Bae, B. H. Hong, Y. H. Ahn, D.-I. Yeom, and F. Rotermund, "Efficient Mode-Locking of Sub-70-fs Ti:Sapphire Laser by Graphene Saturable Absorber," Appl. Phys. Express **5**(3), 032701 (2012).
10. A. Marini, J. D. Cox, and F. J. G. de Abajo, "Theory of graphene saturable absorption," Phys. Rev. B **95**(12), 125408 (2017).
11. H. Zhou, T. Gu, J. F. McMillan, N. Petrone, A. van der Zande, J. C. Hone, M. Yu, G. Lo, D.-L. Kwong, G. Feng, S. Zhou, and C. W. Wong, "Enhanced four-wave mixing in graphene-silicon slow-light photonic crystal waveguides," Appl. Phys. Lett. **105**(9), 091111 (2014).
12. M. Ji, H. Cai, L. Deng, Y. Huang, Q. Huang, J. Xia, Z. Li, J. Yu, and Y. Wang, "Enhanced parametric frequency conversion in a compact silicon-graphene microring resonator," Opt. Express **23**(14), 18679–18685 (2015).
13. K. Alexander, N. A. Savostianova, S. A. Mikhailov, B. Kuyken, and D. Van Thourhout, "Electrically Tunable Optical Nonlinearities in Graphene-Covered SiN Waveguides Characterized by Four-Wave Mixing," ACS Photonics **4**(12), 3039–3044 (2017).
14. N. Kumar, J. Kumar, C. Gerstenkorn, R. Wang, H.-Y. Chiu, A. L. Smirl, and H. Zhao, "Third harmonic generation in graphene and few-layer graphite films," Phys. Rev. B **87**(12), 121406 (2013).
15. S.-Y. Hong, J. I. Dadap, N. Petrone, P.-C. Yeh, J. Hone, and R. M. Osgood Jr., "Optical Third-Harmonic Generation in Graphene," Phys. Rev. X **3**(2), 021014 (2013).
16. A. Ishizawa, R. Kou, T. Goto, T. Tsuchizawa, N. Matsuda, K. Hitachi, T. Nishikawa, K. Yamada, T. Sogawa, and H. Gotoh, "Optical nonlinearity enhancement with graphene-decorated silicon waveguides," Sci. Rep. **7**(1), 45520 (2017).
17. E. Hendry, P. J. Hale, J. Moger, A. K. Savchenko, and S. A. Mikhailov, "Coherent Nonlinear Optical Response of Graphene," Phys. Rev. Lett. **105**(9), 097401 (2010).
18. R. Wu, Y. Zhang, S. Yan, F. Bian, W. Wang, X. Bai, X. Lu, J. Zhao, and E. Wang, "Purely Coherent Nonlinear Optical Response in Solution Dispersions of Graphene Sheets," Nano Lett. **11**(12), 5159–5164 (2011).
19. H. Zhang, S. Virally, Q. Bao, L. K. Ping, S. Massar, N. Godbout, and P. Kockaert, "Z-scan measurement of the nonlinear refractive index of graphene," Opt. Lett. **37**(11), 1856–1858 (2012).
20. W. Chen, G. Wang, S. Qin, C. Wang, J. Fang, J. Qi, X. Zhang, L. Wang, H. Jia, and S. Chang, "The nonlinear optical properties of coupling and decoupling graphene layers," AIP Adv. **3**(4), 042123 (2013).
21. K. Liu, J. F. Zhang, W. Xu, Z. H. Zhu, C. C. Guo, X. J. Li, and S. Q. Qin, "Ultra-fast pulse propagation in nonlinear graphene/silicon ridge waveguide," Sci. Rep. **5**(1), 16734 (2015).
22. G. Demetriou, H. T. Bookey, F. Biancalana, E. Abraham, Y. Wang, W. Ji, and A. K. Kar, "Nonlinear optical properties of multilayer graphene in the infrared," Opt. Express **24**(12), 13033–13043 (2016).
23. E. Dremetsika, B. Dlubak, S.-P. Gorza, C. Ciret, M.-B. Martin, S. Hofmann, P. Seneor, D. Dolfi, S. Massar, P. Emplit, and P. Kockaert, "Measuring the nonlinear refractive index of graphene using the optical Kerr effect method," Opt. Lett. **41**(14), 3281–3284 (2016).
24. N. Vermeulen, D. Castelló-Lurbe, J. Cheng, I. Pasternak, A. Krajewska, T. Ciuk, W. Strupinski, H. Thienpont, and J. Van Erps, "Negative Kerr Nonlinearity of Graphene as seen via Chirped-Pulse-Pumped Self-Phase Modulation," Phys. Rev. Appl. **6**(4), 044006 (2016).
25. N. Vermeulen, D. Castelló-Lurbe, M. Khoder, I. Pasternak, A. Krajewska, T. Ciuk, W. Strupinski, J. Cheng, H. Thienpont, and J. Van Erps, "Graphene's nonlinear-optical physics revealed through exponentially growing self-phase modulation," Nat. Commun. **9**(1), 2675 (2018).
26. J. L. Cheng, N. Vermeulen, and J. E. Sipe, "Third-order nonlinearity of graphene: Effects of phenomenological relaxation and finite temperature," Phys. Rev. B **91**(23), 235320 (2015).
27. S. A. Mikhailov, "Quantum theory of the third-order nonlinear electrodynamic effects of graphene," Phys. Rev. B **93**(8), 085403 (2016).
28. S. A. Mikhailov, "Nonlinear Electrodynamic Properties of Graphene and Other Two-Dimensional Materials," Sens. Transducers **225**, 25–35 (2018).
29. N. A. Savostianova and S. A. Mikhailov, "Optical Kerr effect in graphene: Theoretical analysis of the optical heterodyne detection technique," Phys. Rev. B **97**(16), 165424 (2018).

30. J. L. Cheng, N. Vermeulen, and J. E. Sipe, "Third order optical nonlinearity of graphene," New J. Phys. **16**(5), 053014 (2014).
31. K. Alexander, N. A. Savostianova, S. A. Mikhailov, D. Van Thourhout, and B. Kuyken, "Gate-Tunable Nonlinear Refraction and Absorption in Graphene-Covered Silicon Nitride Waveguides," ACS Photonics **5**(12), 4944–4950 (2018).
32. J. E. Sharping, K. F. Lee, M. A. Foster, A. C. Turner, B. S. Schmidt, M. Lipson, A. L. Gaeta, and P. Kumar, "Generation of correlated photons in nanoscale silicon waveguides," Opt. Express **14**(25), 12388–12393 (2006).
33. C. Xiong, G. D. Marshall, A. Peruzzo, M. Lobino, A. S. Clark, D.-Y. Choi, S. J. Madden, C. M. Natarajan, M. G. Tanner, R. H. Hadfield, S. N. Dorenbos, T. Zijlstra, V. Zwiller, M. G. Thompson, J. G. Rarity, M. J. Steel, B. Luther-Davies, B. J. Eggleton, and J. L. O'Brien, "Generation of correlated photon pairs in a chalcogenide $As_2S_3$ waveguide," Appl. Phys. Lett. **98**(5), 051101 (2011).
34. N. Matsuda, H. Takesue, K. Shimizu, Y. Tokura, E. Kuramochi, and M. Notomi, "Slow light enhanced correlated photon pair generation in photonic-crystal coupled-resonator optical waveguides," Opt. Express **21**(7), 8596–8604 (2013).
35. N. Matsuda, P. Karkus, H. Nishi, T. Tsuchizawa, W. J. Munro, H. Takesue, and K. Yamada, "On-chip generation and demultiplexing of quantum correlated photons using a silicon-silica monolithic photonic integration platform," Opt. Express **22**(19), 22831–22840 (2014).
36. X. Zhang, Y. Zhang, C. Xiong, and B. J. Eggleton, "Correlated photon pair generation in low-loss double-stripe silicon nitride waveguides," J. Opt. **18**(7), 074016 (2016).
37. K. F. Lee, Y. Tian, H. Yang, K. Mustonen, A. Martinez, Q. Dai, E. I. Kauppinen, J. Malowicki, P. Kumar, and Z. Sun, "Photon-Pair Generation with a 100 nm Thick Carbon Nanotube Film," Adv. Mater. **29**(24), 1605978 (2017).
38. L. J. Wang, C. K. Hong, and S. R. Friberg, "Generation of correlated photons via four-wave mixing in optical fibres," J. Opt. B: Quantum Semiclassical Opt. **3**(5), 346–352 (2001).
39. Q. Lin, F. Yaman, and G. P. Agrawal, "Photon-pair generation by four-wave mixing in optical fibers," Opt. Lett. **31**(9), 1286–1288 (2006).
40. Q. Lin and G. P. Agrawal, "Silicon waveguides for creating quantum-correlated photon pairs," Opt. Lett. **31**(21), 3140–3142 (2006).
41. Q. Lin, F. Yaman, and G. P. Agrawal, "Photon-pair generation in optical fibers through four-wave mixing: Role of Raman scattering and pump polarization," Phys. Rev. A **75**(2), 023803 (2007).
42. R. Kou, Y. Hori, T. Tsuchizawa, K. Warabi, Y. Kobayashi, Y. Harada, H. Hibino, T. Yamamoto, H. Nakajima, and K. Yamada, "Ultra-fine metal gate operated graphene optical intensity modulator," Appl. Phys. Lett. **109**(25), 251101 (2016).
43. R. Kou, Y. Kobayashi, K. Warabi, H. Nishi, T. Tsuchizawa, T. Yamamoto, H. Nakajima, and K. Yamada, "Efficient- and Broadband-Coupled Selective Spot-Size Converters With Damage-Free Graphene Integration Process," IEEE Photonics J. **6**(3), 1–9 (2014).
44. C. A. Husko, A. S. Clark, M. J. Collins, A. De Rossi, S. Combrié, G. Lehoucq, I. H. Rey, T. F. Krauss, C. Xiong, and B. J. Eggleton, "Multi-photon absorption limits to heralded single photon sources," Sci. Rep. **3**(1), 3087 (2013).
45. A. D. Bristow, N. Rotenberg, and H. M. Van Driel, "Two-photon absorption and kerr coefficients of silicon for 850–2200 nm," Appl. Phys. Lett. **90**(19), 191104 (2007).
46. S. A. Mikhailov, "Equations of macroscopic electrodynamics for two-dimensional crystals," APL Photonics **4**(3), 034501 (2019).
47. S. Clemmen, A. Perret, J. Safioui, W. Bogaerts, R. Baets, S.-P. Gorza, P. Emplit, and S. Massar, "Low-power inelastic light scattering at small detunings in silicon wire waveguides at telecom wavelengths," J. Opt. Soc. Am. B **29**(8), 1977–1982 (2012).